\documentclass[%
 reprint,
 amsmath,amssymb,
 aps,
pra,
]{revtex4-1}
\usepackage{silence}
\usepackage[utf8]{inputenc}
\usepackage{graphicx}
\usepackage{dcolumn}
\usepackage{bm}
\usepackage{ulem}
\usepackage{xcolor}
\usepackage{lipsum}
\usepackage{hyperref}
\usepackage{float}
\hypersetup{
    colorlinks=true,
    linkcolor=blue,
    filecolor=magenta,      
    urlcolor=black,
    citecolor=blue,
    pdftitle={Overleaf Example},
    pdfpagemode=FullScreen,
    }

\begin{document}

\preprint{APS/123-QED}

\title{Robustness and optimization of N00N-state interferometry}

\author{Romain Dalidet$^1$, Anthony Martin$^1$, Louis Bellando$^1$, Mathieu Bellec$^1$, Nicolas Fabre$^2$, Sébastien Tanzilli$^1$, Laurent Labonté$^1$}
\email{laurent.labonte@univ-cotedazur.fr}
\affiliation{$^1$Université Côte d’Azur, CNRS, Institut de physique de Nice, France \\ 
$^2$Télécom Paris-LTCI, Institut Polytechnique de Paris, 19 Place Marguerite Perey, 91120 Palaiseau, France}

\begin{abstract}
Quantum-enhanced interferometry is often discussed in terms of ideal resources and asymptotic scalings, whereas in practice its performance is set by a delicate interplay between losses, state imbalance, and photon number. We address this interplay in a folded Franson interferometer fed with partially entangled N00N states, treating asymmetric losses and tunable input imbalance on equal footing. From exact detection probabilities we obtain closed-form expressions for the fringe visibility and the Fisher information, and show that these two figures of merit respond very differently to imperfections. In particular, we demonstrate that perfect interference contrast can always be recovered by compensating loss asymmetry with an appropriate input imbalance, while the Fisher information generally peaks at a distinct operating point, reflecting the irreducible trade-off between coherence restoration and signal attenuation. By determining the exact optima and benchmarking against single-photon strategies, we identify the critical loss and minimum entanglement required to maintain a genuine quantum advantage over optimized single-photon strategies under identical loss conditions, and establish their scaling with the photon number $N$. Beyond delineating the fundamental trade-offs between loss, entanglement, and sensitivity, this work establishes a comprehensive theoretical framework that both underpins and extends the experimental demonstration of quantum advantage reported in \cite{dalidet_QA}, providing a unified description of the relevant operating regimes.

 \end{abstract}

\maketitle

\section{Introduction}

Quantum-enhanced interferometry is often presented as a direct route beyond the standard quantum limit (SQL): prepare a nonclassical probe, imprint a phase, and read out an interference fringe with an increased slope \cite{Barbieri2022,Paris2009}. In realistic experiments, however, the achievable performance is not determined by a single imperfection but by a coupled landscape of experimental parameters in which losses, state imbalance, interferometer asymmetries, and photon number jointly shape the available information \cite{Lee2002Rosetta,Demkowicz2015}. In this landscape, some parameters are genuinely critical, as small deviations can abruptly suppress any quantum advantage \cite{Rubin2007,Oh2017}, whereas others can be tuned to partially compensate each other \cite{Lee2009Loss,AtamanMishra2024}. A central question for practical quantum sensors is therefore not whether an ideal N00N state can beat the SQL \cite{Mitchell2004,Nagata2007,Afek2010}, but rather how quantum advantage deforms under realistic imperfections and which experimental degrees of freedom can be exploited to restore it \cite{ThomasPeter2011,Qin2023,Nielsen2023}.

This issue is particularly acute for N00N-state metrology, where N00N states constitute the paradigmatic resource for super-resolution and Heisenberg scaling in lossless interferometry \cite{Lee2002Rosetta,Mitchell2004,Nagata2007,Huver2008,Hiekkamaki2021,Winsten2024}, yet they are notoriously fragile to loss and decoherence \cite{Rubin2007,Demkowicz2009,Kolodynski2010,Datta2011}. Fundamental bounds have established that in the presence of noise the Heisenberg scaling cannot be maintained and that the achievable precision is ultimately limited to, at best, a constant-factor improvement over the SQL \cite{Dorner2009,Demkowicz2009,Escher2011,Datta2011,ThomasPeter2011}. This fragility has stimulated the search for more robust probes and control strategies \cite{holland_interferometric_1993, Huver2008,Thekkadath2020,Qin2023,Nielsen2023}, as well as the optimization of imperfect interferometers by deliberately exploiting asymmetries, for instance through unbalanced beam splitters or tailored input states, in order to maximize the (quantum) Fisher information in lossy and unbalanced settings \cite{Lee2009Loss,AtamanMishra2024,HuLu2025}.

A crucial point, often overlooked in experimental practice, is that high fringe visibility does not necessarily imply high metrological sensitivity \cite{Braunstein1994,Datta2011,Oh2017}. Visibility is governed by the coherence between the two interferometric paths and can, in many situations, be restored by compensating amplitude imbalance \cite{Lee2009Loss,Agarwal2010}. The Fisher information, by contrast, depends not only on the slope of the interference signal, but also on the absolute detection probabilities, and is therefore sensitive to the conversion of relative losses into an effective global attenuation \cite{Demkowicz2009,Kolodynski2010,AtamanMishra2024}. As a result, the operating point that maximizes contrast does not, in general, coincide with the one that maximizes information. This distinction becomes even more relevant in a multiparameter perspective, where phase and loss (or phase and visibility) are not necessarily compatible parameters and where probe and measurement optimization for one quantity may be suboptimal for the other \cite{Bezerra2025,HuertaAlderete2022}.

In this work we present an analytical, experimentally relevant study of these trade-offs for a Mach-Zehnder-like folded Franson interferometer fed with partially entangled N00N states, parametrized by a tunable input imbalance. We focus on two dominant and experimentally relevant imperfections: asymmetric losses between the interferometer arms and non-maximal path entanglement, equivalently an unbalanced input beam splitter \cite{Lee2009Loss,AtamanMishra2024}. From exact detection probabilities we derive closed-form expressions for the $N$-photon fringe visibility and for the (normalized) Fisher information. We show that losses and input imbalance can compensate each other at the level of interference contrast: by appropriately tuning the entanglement, perfect visibility can be restored even in the presence of strong asymmetrical losses \cite{Agarwal2010,Lee2009Loss}. However, this compensation does not, in general, maximize the Fisher information \cite{Demkowicz2009,AtamanMishra2024}. The operating point that optimizes contrast differs from that which optimizes sensitivity, reflecting the fact that restoring coherence does not compensate for the information loss associated with reduced detection rates. By benchmarking against single-photon strategies under identical loss conditions \cite{ThomasPeter2011,Oh2017}, we identify the parameter regimes in which multi-photon N00N interference still provides a genuine quantum advantage \cite{Qin2023,Nielsen2023,Thekkadath2020,Wang2024}, and we delineate which experimental parameters act as critical limitations and which can be used as soft control knobs for performance optimization.

\section{Model}\label{SECTION 2}

\begin{figure}[ht]
    \centering
    \includegraphics[width=1\linewidth]{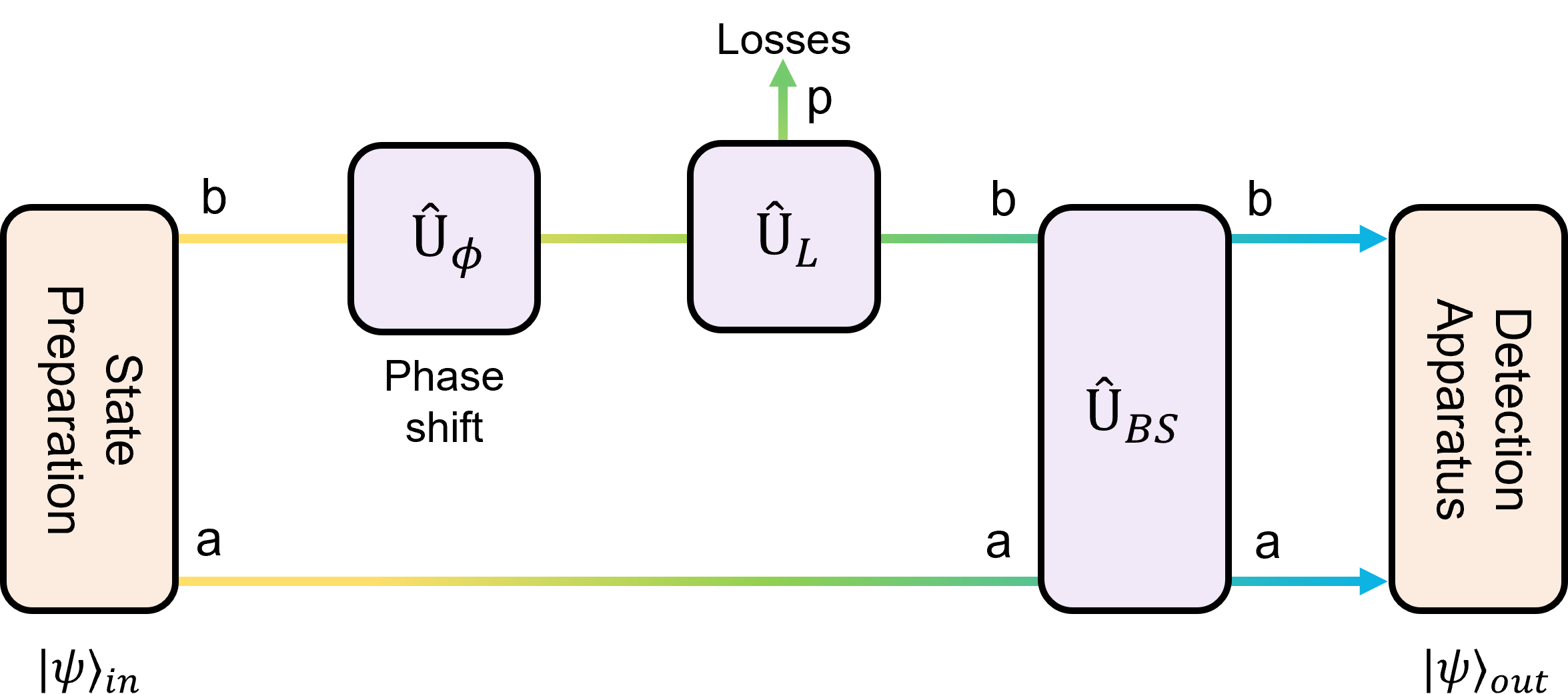}
   \caption{General scheme of the interferometric configuration considered in this work, based on a N00N state propagating in a Mach--Zehnder--like folded Franson interferometer. The two arrows represent the lower and upper arms of the interferometer. A relative phase shift and relative losses are introduced in the upper arm and are modeled by the unitary transformations $\hat{U}_{\phi}$ and $\hat{U}_{L}$, respectively.}

    \label{fig: theoretical scheme}
\end{figure}

The optical configuration considered in this work is shown in Fig.~\ref{fig: theoretical scheme} and corresponds to a folded Franson interferometer \cite{Franson1989,Rarity1990}. The input state is generated in the "State Preparation" stage and consists of a partially entangled path-encoded $N$-photon state of the form
\begin{equation}\label{eq: psi init}
    |\psi^{\mathrm{in}}_{N}\rangle_{ab}
    = \Bigl[\sqrt{\alpha}\,(\hat a^\dagger)^N
    + \sqrt{1-\alpha}\,(\hat b^\dagger)^N \Bigr] |00\rangle_{ab},
\end{equation}
where $N$ denotes the total photon number and the modes $a$ and $b$ label the lower and upper arms of the interferometer, respectively. The parameter $\alpha\in[0,1]$ controls the degree of path entanglement, with $\alpha=1/2$ corresponding to a maximally entangled N00N state. Throughout this work we assume the input state to be pure \textit{i.e.} $\text{Tr}(\rho^2 )=1$ where $\rho$ is the density matrix of the input state. Note that for $N=1$ the state reduces to a nonclassical superposition of the two interferometer paths, which we use as a reference case throughout this work.

\begin{figure*}[!t]
    \centering
    \includegraphics[width=0.65\linewidth]{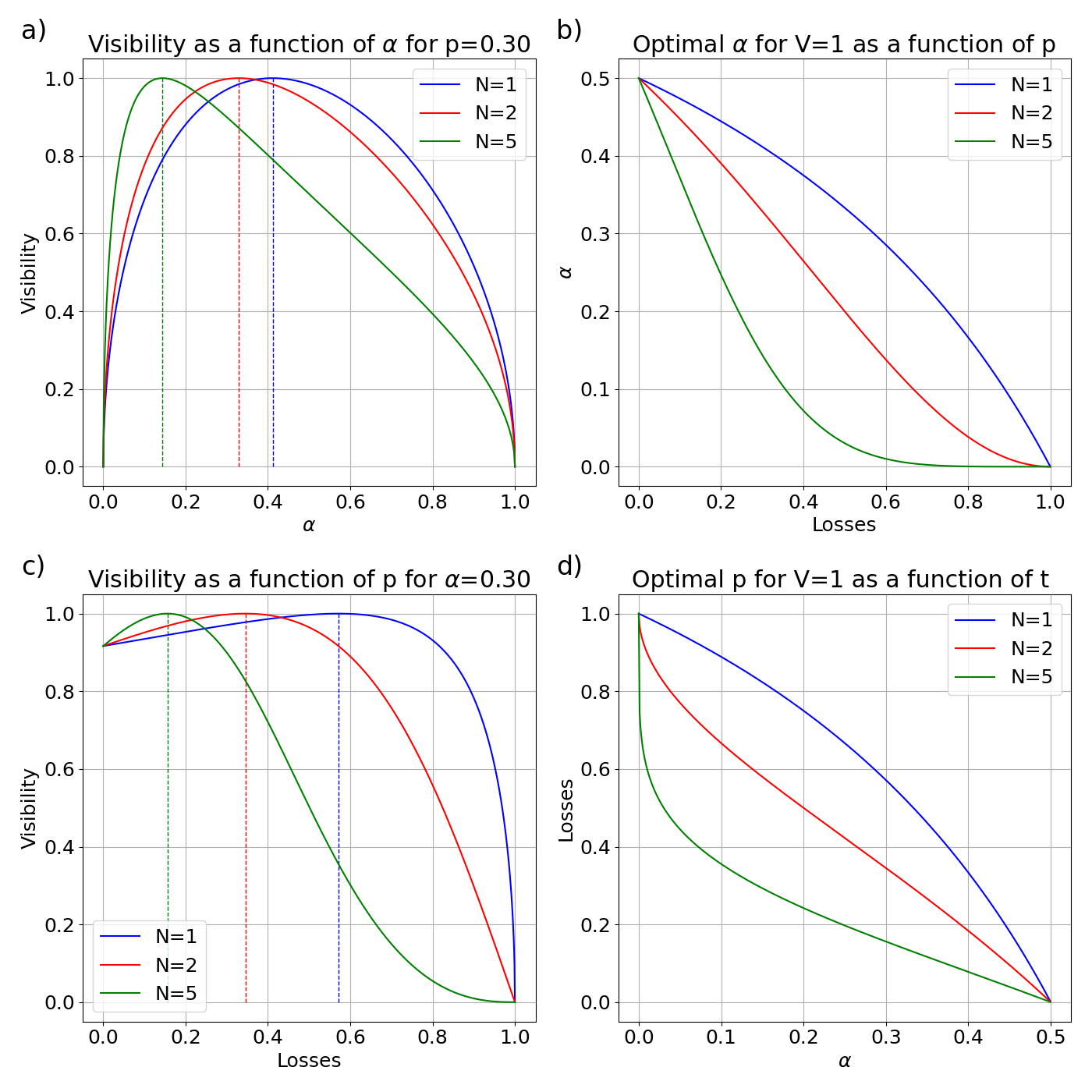}
  \caption{(a) Fringe visibility [Eq.~(\ref{eq: visibility})] as a function of the entanglement parameter $\alpha$ for fixed relative losses ($p=0.3$), for $N=1,2,$ and $5$ photons (blue, red, and green curves, respectively). Dotted lines indicate the global maxima, showing that the visibility can be fully restored by an appropriate tuning of $\alpha$. (b) Optimal entanglement parameter $\alpha_{v}$ given by Eq.~(\ref{eq: Vis opt t}) as a function of the relative losses, yielding the maximum visibility; in the lossless limit the optimum corresponds to a maximally entangled state.  (c) Visibility as a function of the relative losses $p$ for a fixed degree of entanglement ($\alpha=0.3$). Dotted lines mark the local maxima. Increasing the photon number makes the visibility increasingly sensitive to losses.  (d) Optimal relative loss $p_{v}$ given by Eq.~(\ref{eq: Vis opt p}) as a function of the entanglement parameter, maximizing the visibility.}

    \label{fig: etude visibilité}
\end{figure*}
In interferometric phase measurements, one arm is typically used as a reference, while the other contains the sample or process to be probed. In our configuration, the lower arm plays the role of a stable reference with fixed phase and negligible losses, whereas the upper arm accumulates both a relative phase shift and relative losses. These two effects are modeled by the unitary transformations
\begin{equation}
\hat{b}^{\dagger} \xrightarrow{\hat{U}_\phi} e^{i\phi}\hat{b}^{\dagger},
\qquad
\hat{b}^{\dagger} \xrightarrow{\hat{U}_L} \sqrt{p}\,\hat{p}^{\dagger} + \sqrt{1-p}\,\hat{b}^{\dagger}.
\label{eq: uni phi losses}
\end{equation}

where $\phi$ is the phase to be estimated and $p$ denotes the loss probability in the upper arm, so that $1-p$ is the corresponding transmission. Losses are modeled by a fictitious unbalanced beam splitter coupling the interferometric mode $\hat b$ to an environmental vacuum mode $\hat p$. When both arms exhibit losses, the problem can always be recast in terms of relative losses by factoring out a global attenuation from the input state.

The two arms are then recombined on a balanced beam splitter, where the interference between the two $N$-photon amplitudes takes place. Since the measurement consists of $N$-photon coincidence counts at the two output ports, the environmental mode associated with losses is traced out. The resulting reduced output state of the interferometer can therefore be written in the form
\begin{equation}\label{eq: psi out}
    |\psi^{\mathrm{out}}_{N}\rangle_{ab}
    = \Bigl[ C_{N}(\hat{a}^{\dagger}+\hat{b}^{\dagger})^{N}
    + D_{N} e^{iN\phi}(\hat{a}^{\dagger}-\hat{b}^{\dagger})^{N} \Bigr] |00\rangle_{ab},
\end{equation}
where the coefficients
\begin{equation}\label{eq: Cn and Dn}
    C_{N} = \frac{\sqrt{\alpha}}{2^{N/2}}, \qquad
    D_{N} = \frac{\sqrt{1-\alpha}\,(1-p)^{N/2}}{2^{N/2}}
\end{equation}
encode, respectively, the weight of the reference-arm amplitude and the attenuated, phase-shifted amplitude propagating through the lossy arm. Equation~(\ref{eq: psi out}) thus describes a generalized N00N state at the interferometer output, in which the relative phase $\phi$ and the relative losses $p$ appear only through the relative phase factor $e^{iN\phi}$ and the amplitude imbalance between $C_N$ and $D_N$.

In the present model we neglect global losses such as imperfect beam splitters or finite detector efficiencies. Such effects would simply introduce an overall attenuation factor multiplying Eq.~(\ref{eq: psi out}) \cite{Rubin2007} and do not affect the relative trade-offs discussed below. Equation~(\ref{eq: psi out}) thus describes a generalized N00N state of arbitrary path entanglement propagating through a lossy interferometer. Finally, since the loss and phase operations commute, $\left[\hat{U}_{L},\hat{U}_{\phi}\right]=0$, the phase shift may be equivalently applied before or after the fictitious beam splitter modeling the losses \cite{Huver2008}.

From the output state derived above, we now turn to the observable quantities that characterize the interference pattern and its metrological content. The measurement consists of $N$-photon coincidence detection at the two output ports of the interferometer. The corresponding joint probability to register $i$ photons in mode $a$ and $j=N-i$ photons in mode $b$ is obtained from the output state $|\psi^{\mathrm{out}}_{N}\rangle_{ab}$ and can be written in the closed form

\begin{equation}\label{eq: detection}
P_{ij}
= \binom{N}{i}\, A_N \Bigl[1+(-1)^j V_N \cos(N\phi)\Bigr],
\end{equation}
where the binomial prefactor reflects the indistinguishability of the $N$ photons and the binomial structure of the output amplitudes. The quantities
\begin{equation}
A_N = C_N^2 + D_N^2 , \qquad 
V_N = \frac{2C_ND_N}{C_N^2+D_N^2}
\end{equation}
set, respectively, the overall weight of the interference pattern and its visibility. Importantly, both are independent of the specific output partition $(i,j)$ and depend only on the relative amplitudes of the two $N$-photon paths.

Summing over all coincidence outcomes yields
\begin{equation}\label{eq: energy conservation}
\sum_{i=0}^{N} P_{i,N-i}
= \alpha + (1-\alpha)(1-p)^N ,
\end{equation}
which reduces to unity in the absence of relative losses and, in that case, becomes independent of the entanglement parameter. This result directly reflects the conversion of relative losses into a global attenuation factor for the $N$-photon component.

In the following, we first exploit these analytical expressions to study how the fringe visibility depends on the entanglement parameter $\alpha$, the relative losses $p$, and the photon number $N$, and to identify the operating points that maximize contrast. We then address the metrological performance by analyzing the corresponding Fisher information, thereby determining the regimes in which the restoration of interference does or does not translate into an optimal phase sensitivity, and establishing the conditions for a genuine quantum advantage over single-photon strategies.

\section{Performance analysis}\label{SECTION 3}
\subsection{Fringes visibility}

The visibility of the $N$-photon interference fringes follows directly from the detection probabilities and reads
\begin{equation}\label{eq: visibility}
    V(\alpha,p,N)
    = 2\,\frac{\sqrt{\alpha(1-\alpha)}\,(1-p)^{N/2}}
           {\alpha+(1-\alpha)(1-p)^N}.
\end{equation}
For all physical values $\alpha,p\in[0,1]$ and arbitrary $N$, one has $0\leq V\leq1$. In the absence of relative losses ($p=0$), the visibility becomes independent of the photon number and is solely determined by the degree of entanglement. Representative behaviors as a function of $\alpha$ and $p$ are shown in Figs.~\ref{fig: etude visibilité}(a) and \ref{fig: etude visibilité}(c).

The extrema of Eq.~(\ref{eq: visibility}) with respect to $\alpha$ and $p$ reveal a single global maximum, obtained from
\begin{align}
    \frac{\partial V}{\partial \alpha}=0
    &\;\Longleftrightarrow\;
    \alpha_{\!V}(p,N)
    = \frac{(1-p)^N}{1+(1-p)^N}, \label{eq: Vis opt t}\\
    \frac{\partial V}{\partial p}=0
    &\;\Longleftrightarrow\;
    p_{\!V}(\alpha,N)
    = 1-\!\left(\frac{\alpha}{1-\alpha}\right)^{\!1/N}. \label{eq: Vis opt p}
\end{align}
Equation~(\ref{eq: Vis opt t}) gives the optimal degree of entanglement that maximizes the fringe contrast for given losses and photon number. In the lossless limit, it reduces to $\alpha=1/2$, corresponding to a maximally entangled N00N state, independently of $N$. In the opposite regime of strong attenuation, the optimum shifts toward increasingly unbalanced superpositions, as illustrated in Fig.~\ref{fig: etude visibilité}(b).

Remarkably, inserting Eq.~(\ref{eq: Vis opt t}) into Eq.~(\ref{eq: visibility}) yields $V=1$ for any $p$ and $N$. Perfect fringe contrast can therefore always be restored by appropriately matching the degree of entanglement to the loss asymmetry. At the state level, this compensation condition leads to
\begin{equation}
\begin{split}
|\psi^{\mathrm{out}}_N\rangle_{ab}
= {} & \frac{(1-p)^{N/2}}{2^{N/2}\sqrt{1+(1-p)^N}} \\
& \times
\Big[
(\hat a^\dagger + \hat b^\dagger)^N
+ e^{iN\phi}(\hat a^\dagger - \hat b^\dagger)^N
\Big]
|00\rangle_{ab} .
\end{split}
\end{equation}
showing that relative losses are converted into a global attenuation factor that affects only the overall amplitude, while leaving the interference contrast unchanged.

Equation~(\ref{eq: Vis opt p}), which is simply the inverse of Eq.~(\ref{eq: Vis opt t}), provides the loss value that maximizes visibility for a given entanglement and photon number, as displayed in Fig.~\ref{fig: etude visibilité}(d). These results demonstrate that, at the level of fringe contrast, loss and state imbalance act as fully compensable parameters: for any $N$, one can trade relative attenuation against entanglement asymmetry to recover unit visibility. As discussed in the next section, this compensation mechanism, although sufficient to restore coherence, does not in general coincide with the operating point that maximizes the metrological information.

\subsection{Fisher information}

To ascertain the sensitivity achievable in an interferometric experiment, we employ the Fisher information (FI), which quantifies the amount of information that can be extracted per photon transmitted through the system \textit{i.e.} the minimum phase variation that can be measured. Unlike visibility, FI directly relates to the achievable precision of phase estimation and can behave differently under loss and imperfect entanglement. In this context, we define the normalized FI as \cite{Braunstein1994}:

\begin{equation}\label{eq: Fisher info 1}
\begin{split}
\mathcal{F}_N &= \frac{1}{N}\sum_{\substack{i=0 \\ j=N-i}}^{N} 
\left( \frac{\partial \ln P_{ij}}{\partial \phi} \right)^2 P_{ij} \\
&= \frac{1}{N}\sum_{\substack{i=0 \\ j=N-i}}^{N} 
\frac{(P_{ij}')^2}{P_{ij}} .
\end{split}
\end{equation}
It is essential to highlight the presence of the factor $1/N$, which normalizes the FI by the number of entangled photons. This choice ensures that the standard quantum limit (SQL) corresponds to $\mathcal{F}_N = 1$ for any N, allowing direct comparison between single-photon and N-photon strategies. Combining the previous equation with the detection probabilities defined by Eq. \ref{eq: detection}, we obtain:

\begin{equation}\label{eq: Fisher info 2}
    \mathcal{F}_N = N 2^N \frac{A_N V_N^2 sin^2(N\phi)}{1-V_Ncos^2(N\phi)}\, .
\end{equation}
The last equation shows that FI depends on both the amplitude and visibility of the interference pattern, as well as its slope defined by the derivative term. Fig. \ref{fig:  proba et FI} a) and b) illustrate the probability of detection and associated FI for single-photon and two-photon interference, respectively. The FI reaches its maximum value when $\phi = \frac{\pi}{2N}(2k+1)$ where $k \in \mathbb{N}$, that is, where the slope of the interference pattern is the steepest. In this case we obtain:

\begin{figure}[t!]
    \centering
    \includegraphics[width=1\linewidth]{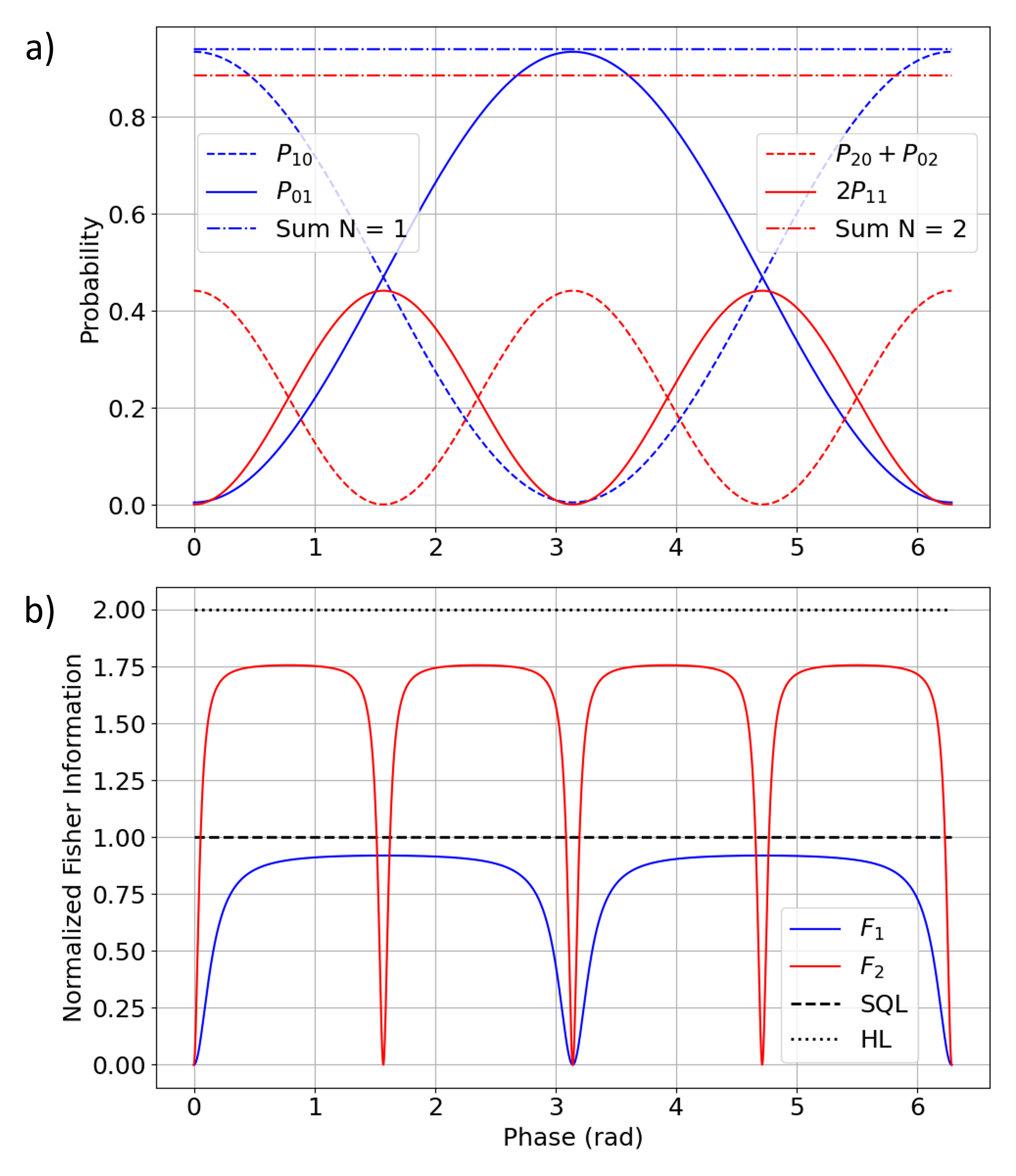}
    \caption{(a) Output detection probabilities for $N=1$ (blue) and $N=2$ (red) photons, for relative loss $p=0.1$ and entanglement parameter $\alpha=0.4$. The dotted curves correspond to the total probability given by Eq.~(\ref{eq: energy conservation}). 
(b) Corresponding Fisher information computed from Eq.~(\ref{eq: Fisher info 2}). The lower and upper dotted horizontal lines indicate the standard quantum limit and the Heisenberg limit, respectively.}
    \label{fig:  proba et FI}
\end{figure}

\begin{equation}\label{eq: Fisher info max}
\begin{split}
    \mathcal{F}_N  & = N2^N A_N V_N^2 \\
                   & = 4N \frac{(\alpha - \alpha^2)(1-p)^N}{\alpha+(1-\alpha)(1-p)^N} \, .
\end{split}
\end{equation}
\begin{figure*}[!]
    \centering
    \includegraphics[width=1\linewidth]{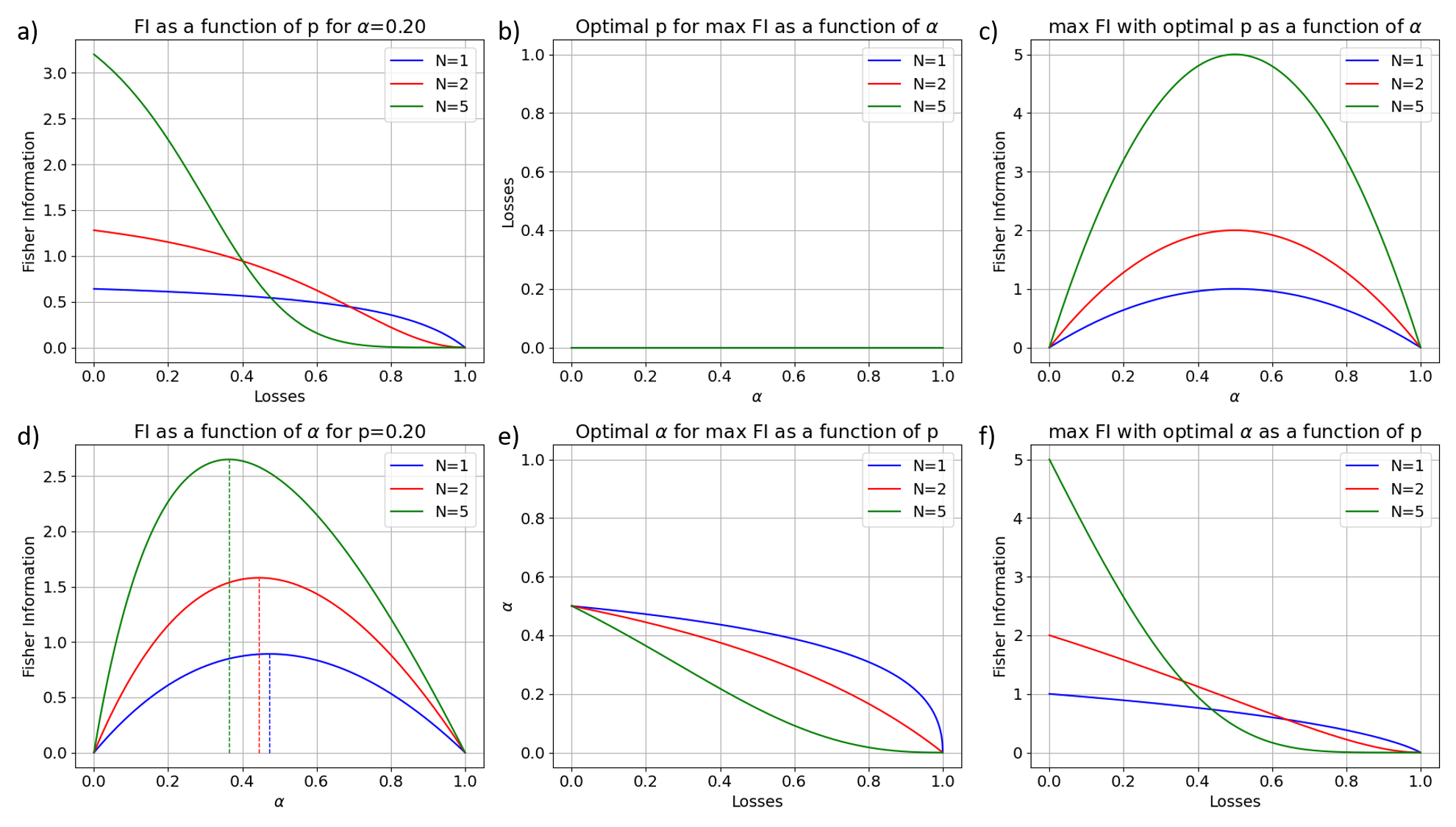}
   \caption{(a) Fisher information [Eq.~(\ref{eq: Fisher info max})] as a function of the relative loss $p$ for a fixed entanglement parameter $\alpha=0.2$ and $N=1,2,$ and $5$ photons (blue, green, and red curves, respectively). At low loss, increasing $N$ enhances the sensitivity, but also makes it more fragile to attenuation.  (b) Optimal relative loss maximizing the Fisher information [Eq.~(\ref{eq: FI opt p})], showing that for any $\alpha$ and $N$ the maximum is always reached in the lossless limit.  (c) Maximum Fisher information at optimal loss [Eq.~(\ref{eq: FI max opt p})] as a function of $\alpha$, exhibiting a peak at $\alpha=1/2$ and a linear scaling with $N$.  (d) Fisher information as a function of $\alpha$ for $p=0.2$; dotted lines indicate the optimal values.  (e) Optimal entanglement parameter $\alpha_{\mathrm{opt}}$ maximizing the Fisher information [Eq.~(\ref{eq: FI opt t})] as a function of the relative loss. The optimum shifts toward increasingly unbalanced states as $p$ increases.  (f) Maximum Fisher information at optimal entanglement [Eq.~(\ref{eq: FI max opt t})] as a function of $p$. The sensitivity decreases monotonically with loss, with a faster degradation for larger photon numbers.}
\label{fig: etude FI p et t}
\end{figure*}
Thus, $F_N \neq V_N$, optimizing the visibility of the interference fringes does not optimize the FI. As previously stated, it is always possible to enhance the visibility of the interference, which will inevitably result in a multiplicative factor influencing the final amplitude of the fringes. Given that the FI is determined by the slope of the interference fringes, a decrease in amplitude will consequently affect it. In the ideal scenario where $\alpha = 0.5$ and $p = 0$ (lossless system with a maximally entangled input quantum state), we have $V_N = 1$ and $F_N = N$. The latter result defines the SQL or the Heisenberg limit for single and N-photon interference, respectively. In the following sub-sections, we examine Eq. \ref{eq: Fisher info max} as a function of $\alpha$ and $p$ to optimize the maximum value of the FI reachable.

\subsubsection{Fisher-information optimization} \label{SECTION 3A}

We now turn to the optimization of the phase sensitivity, quantified by the Fisher information $\mathcal{F}_N$. As a first step, we examine its dependence on the relative losses. Taking the derivative of Eq.~(\ref{eq: Fisher info max}) with respect to $p$ yields
\begin{equation}\label{eq: deriv FI p}
\frac{\partial \mathcal{F}_N}{\partial p}
= 4N^2\,\frac{\alpha^2(\alpha-1)(1-p)^N}
{\bigl[\alpha+(1-\alpha)(1-p)^N\bigr]^2}.
\end{equation}
The prefactor $(\alpha-1)$ shows that this derivative is always negative for any $0<\alpha<1$, implying that the Fisher information decreases monotonically with increasing loss, independently of the photon number and of the degree of entanglement, in agreement with general bounds for lossy interferometry \cite{Demkowicz2009}. As a consequence, there is no nontrivial optimum in $p$:
\begin{equation}\label{eq: FI opt p}
p_{\mathrm{opt}}(\alpha,N)=0,
\end{equation}
and the maximal Fisher information at fixed $\alpha$ and $N$ is reached in the lossless limit,
\begin{equation}\label{eq: FI max opt p}
\mathcal{F}_N(\alpha,p_{\mathrm{opt}},N)
=4N(\alpha-\alpha^2),
\end{equation}
which itself is maximized for a balanced, maximally entangled input state. Representative behaviors are shown in Figs.~\ref{fig: etude FI p et t}(a)--(c). In the absence of relative losses, increasing the photon number always enhances the sensitivity, whereas this monotonic scaling is rapidly suppressed as soon as $p>0$, as discussed below.

We next optimize the Fisher information with respect to the degree of entanglement. As a function of $\alpha$, Eq.~(\ref{eq: Fisher info max}) is concave and admits a single maximum, obtained from
\begin{equation}\label{eq: deriv FI t}
\frac{\partial \mathcal{F}_N}{\partial \alpha}
=4N(1-p)^N\,
\frac{(1-\alpha)^2(1-p)^N-\alpha^2}
{\bigl[(1-\alpha)(1-p)^N+\alpha\bigr]^2},
\end{equation}
which vanishes for
\begin{equation}\label{eq: FI opt t}
\alpha_{\mathrm{opt}}(p,N)
=\frac{(1-p)^{N/2}}{1+(1-p)^{N/2}}.
\end{equation}
Inserting this optimal value into Eq.~(\ref{eq: Fisher info max}) yields
\begin{equation}\label{eq: FI max opt t}
\mathcal{F}_N(\alpha_{\mathrm{opt}},p,N)
=4N\,\frac{(1-p)^N}{\bigl[1+(1-p)^{N/2}\bigr]^2},
\end{equation}
which makes explicit how the ultimate sensitivity is progressively degraded by relative losses, and reaches its maximum in the ideal, lossless limit. Examples of these optimal behaviors are shown in Figs.~\ref{fig: etude FI p et t}(d)--(f).

A striking parallel emerges when comparing the optimal entanglement for visibility and for Fisher information. Equations~(\ref{eq: Vis opt t}) and (\ref{eq: FI opt t}) are formally related through
\[
\alpha_V(p,N)=\alpha_{\mathrm{opt}}(p,2N),
\]
that is, the degree of entanglement that restores unit contrast for $N$-photon interference coincides with the one that maximizes the Fisher information for a $2N$-photon state. This highlights that, although loss and state imbalance can be perfectly compensated at the level of fringe visibility, the operating point that optimizes metrological information follows a different scaling and reflects a fundamentally distinct trade-off between coherence and attenuation.

\subsubsection{Quantum superiority and quantum advantage}\label{SECTION 3B}

Using the normalized Fisher information defined in Eq.~(\ref{eq: Fisher info 1}), we distinguish two operational notions. We first define quantum superiority as the regime in which the phase sensitivity surpasses the standard quantum limit (SQL),
\begin{equation}\label{eq: def superiorité quantique}
    \mathcal{F}_N(\alpha,p,N) > 1 ,
\end{equation}
which necessarily requires $N>1$. More generally, we define quantum advantage as the situation where entangled probes outperform single-photon probes under identical experimental conditions,
\begin{equation}\label{eq: def adv quantique}
    \frac{\mathcal{F}_N(\alpha,p,N)}{\mathcal{F}_1(\alpha,p,1)} > 1 ,
\end{equation}
even if the SQL is not exceeded. Quantum superiority thus captures a scaling advantage with respect to classical resources, while quantum advantage quantifies the practical gain over the best single-photon strategy in the same lossy environment.

\begin{figure}[h]
    \centering
    \includegraphics[width=1\linewidth]{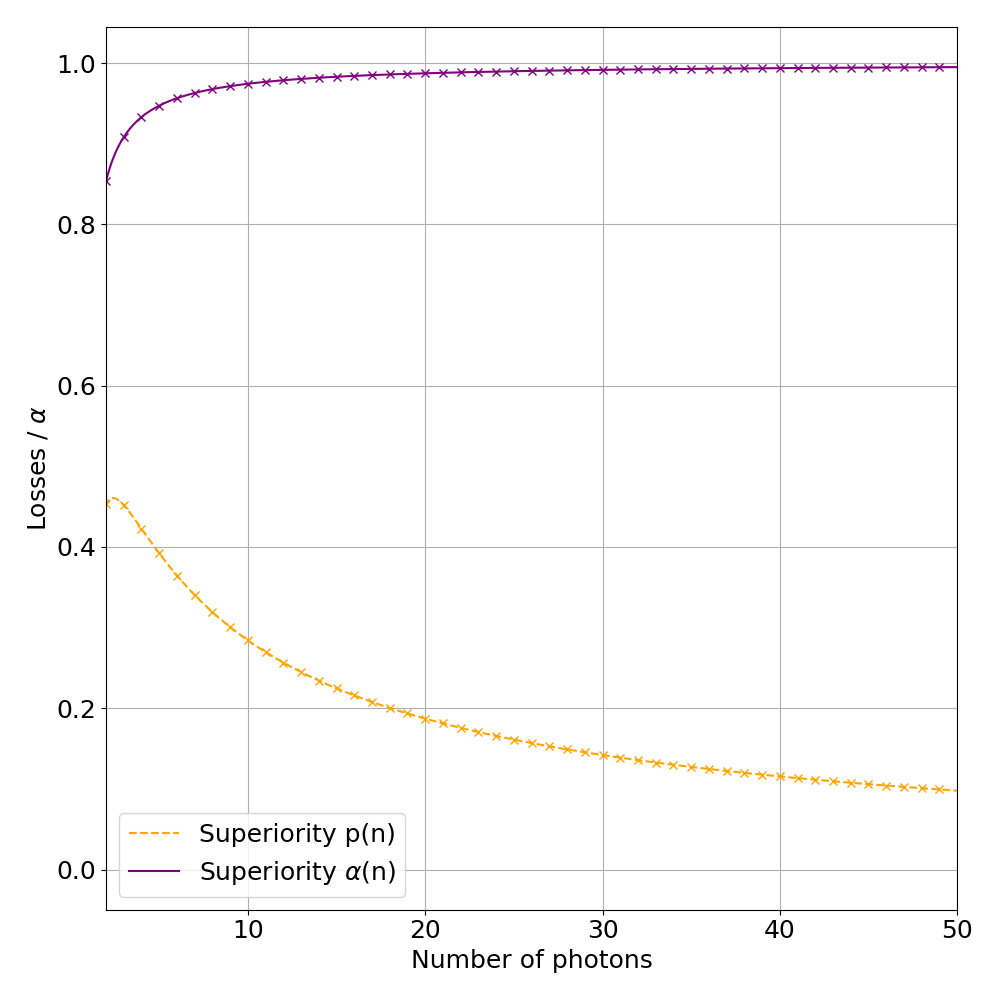}
    \caption{ Boundaries for surpassing the SQL with optimized parameters as a function of $N$. Purple curve shows the minimum degree of entanglement from Eq. \ref{eq: sup opt p} at zero relative losses. The orange curve shows the maximum tolerable relative loss from Eq. \ref{eq: sup opt t} when $\alpha$ is optimized. As $N$ increases, the tolerance to imperfect entanglement increase while the maximum acceptable relative losses decreases.}
    \label{fig:  superiority}
\end{figure}
\begin{figure*}[!t]
    \centering
    \includegraphics[width=0.7\linewidth]{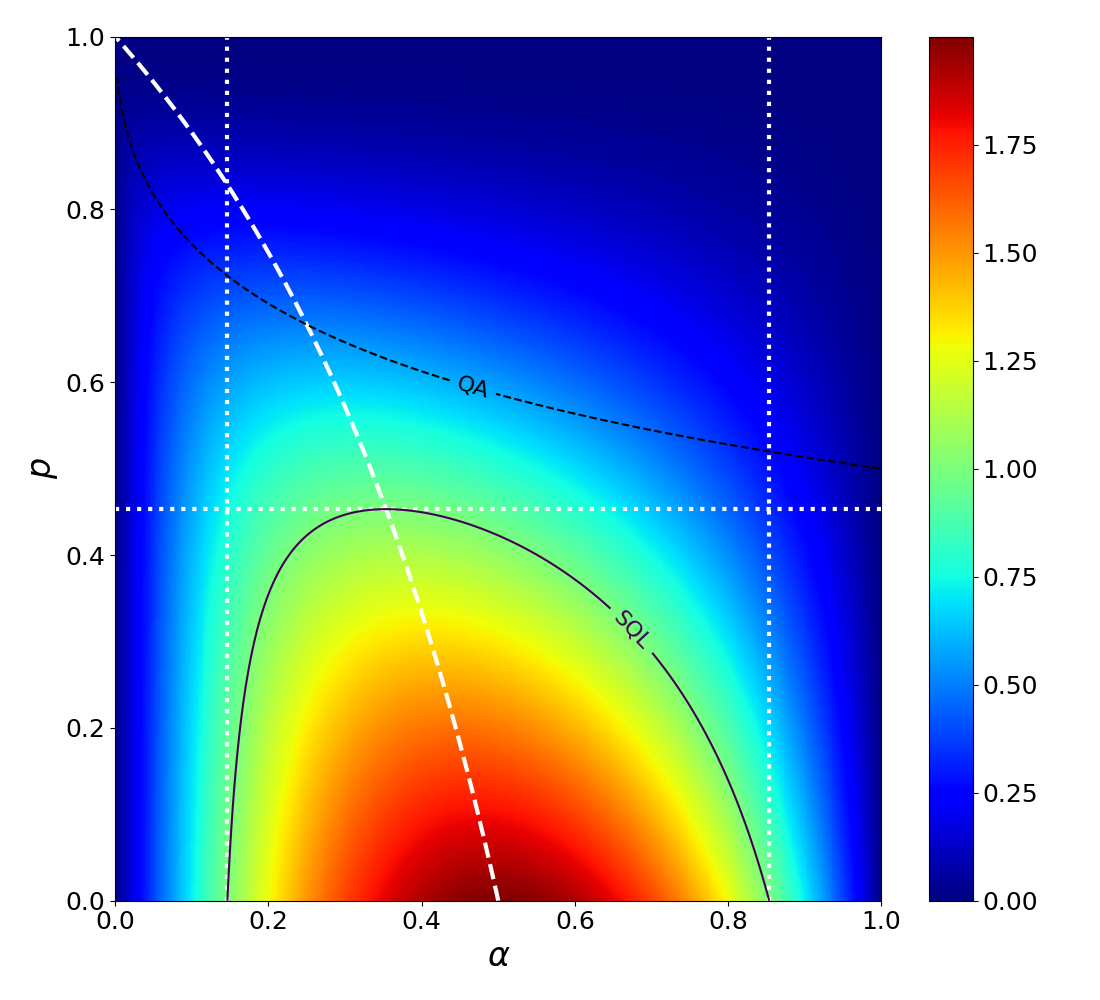}
    \caption{Fisher information for $N=2$ entangled photons as a function of the relative losses $p$ and the entanglement parameter $\alpha$. The Heisenberg limit is attained only in the ideal case of vanishing relative losses and a maximally entangled input state. The SQL boundary, defining the onset of quantum superiority, is given by Eq.~(\ref{eq: superiority p}), while the region of quantum advantage with respect to single-photon interferometry is determined by Eq.~(\ref{eq: adv p}). The two vertical white lines indicate the minimal entanglement required to surpass the SQL in the lossless limit [$p=0$, Eq.~(\ref{eq: sup opt p})]. The horizontal white line marks the maximum relative loss compatible with quantum superiority when the entanglement is optimized [Eq.~(\ref{eq: sup opt t})]. The solid white curve shows the optimal entanglement $\alpha_{\mathrm{opt}}(p,2)$ that maximizes the Fisher information [Eq.~(\ref{eq: FI opt t})]. Its intersection with the quantum-advantage boundary defines the threshold $p\simeq0.64$ [Eq.~(\ref{eq: adv opt t})], beyond which entangled-photon probes no longer outperform single-photon strategies.}

    \label{fig:  final N2}
\end{figure*}

Both criteria lead to analytical boundaries in the $(\alpha,p)$ parameter space. For quantum superiority, one finds
\begin{equation}\label{eq: superiority p}
    p< 1 - \left(\frac{\alpha}{(1-\alpha)(4\alpha N-1)}\right)^{1/N},
\end{equation}
which can be inverted to obtain the corresponding minimal entanglement for a given loss and photon number. For quantum advantage, the condition reads
\begin{equation}\label{eq: adv p}
    N(1-p)^{N-1}\frac{1+p(\alpha-1)}{\alpha+(1-\alpha)(1-p)^N}>1 \, ,
\end{equation}
whose explicit form defines the region where multi-photon interference remains beneficial even when the SQL is not beaten.

Particularly transparent bounds are obtained when the parameters are optimized. In the lossless case ($p=0$), quantum superiority is achieved whenever
\begin{equation}\label{eq: sup opt p}
    \alpha \in \left[\frac{1}{2}\pm\frac{\sqrt{N^2-N}}{2N}\right],
\end{equation}
as shown in Fig.~\ref{fig: superiority} (purple curve). As $N$ increases, the admissible deviation from maximal entanglement widens, indicating that the minimal entanglement required to beat the SQL decreases with photon number. In the same ideal limit, the quantum advantage condition reduces to
\begin{equation}\label{eq: adv opt p}
    \frac{\mathcal{F}_N(\alpha,0,N)}{\mathcal{F}_1(\alpha,0,1)} = N ,
\end{equation}
so that entangled probes always outperform single photons, independently of $\alpha$.

When the degree of entanglement is optimized, $\alpha=\alpha_{\mathrm{opt}}(p,N)$, the condition for quantum superiority becomes
\begin{equation}\label{eq: sup opt t}
    p < 1 - \left(\frac{1+2\sqrt{N}}{4N-1}\right)^{2/N},
\end{equation}
plotted in Fig.~\ref{fig: superiority} (orange curve). In contrast with the lossless case, the maximum tolerable relative loss now decreases with $N$, reflecting the increasing fragility of N00N states: although larger photon numbers relax the entanglement requirements, they tighten the loss budget needed to maintain sub-SQL sensitivity.

A similar analysis yields the boundary for quantum advantage at optimal entanglement,
\begin{equation}
    \frac{\mathcal{F}_N(\alpha_{\mathrm{opt}},p,N)}
         {\mathcal{F}_1(\alpha_{\mathrm{opt}},p,1)}
    = \frac{N(1+\sqrt{1-p})^2(1-p)^{N-1}}
           {\bigl[1+(1-p)^{N/2}\bigr]^2} > 1 ,
\end{equation}
which can be solved analytically for small $N$. For the experimentally relevant two-photon case, this yields
\begin{equation}\label{eq: adv opt t}
    p < \sqrt{31+22\sqrt{2}}-3(1+\sqrt{2}) \simeq 0.64 ,
\end{equation}
providing a concrete benchmark: with photon pairs, a genuine quantum advantage over single-photon interferometry survives up to relative losses of about $64\%$.

\subsubsection{Two-photon case ($N=2$)}

Figure~\ref{fig: final N2} provides a compact summary of these regimes for the experimentally relevant case $N=2$ in the $(\alpha,p)$ plane~\cite{dalidet_QA}. The vertical white lines mark the minimal entanglement required to surpass the SQL in the absence of relative losses [Eq.~(\ref{eq: sup opt p})], while the horizontal white line gives the maximum loss compatible with quantum superiority when the entanglement is optimized [Eq.~(\ref{eq: sup opt t})]. The dotted curve corresponds to the optimal entanglement $\alpha_{\mathrm{opt}}(p,2)$ that maximizes the Fisher information, and its intersection with the quantum-advantage boundary defines the threshold $p\simeq0.64$ of Eq.~(\ref{eq: adv opt t}). Together, these boundaries provide an operational map of the parameter space, highlighting how loss and entanglement jointly constrain the attainable sensitivity and delineating the regions of sub-SQL sensitivity (quantum superiority) and of practical gain over single-photon strategies (quantum advantage), including regimes where advantage persists despite $\mathcal{F}_N < 1$.

\section{Conclusion}\label{SECTION 6}

We have investigated how asymmetric losses and non-maximal entanglement jointly shape the performance of N00N-state interferometry in a folded Mach--Zehnder--like Franson configuration. By deriving closed-form expressions for both the fringe visibility and the Fisher information, we have shown that these two figures of merit respond in fundamentally different ways to imperfections. While relative losses can always be compensated by an appropriate tuning of the input imbalance so as to restore unit interference contrast, this compensation does not, in general, coincide with the operating point that maximizes the available phase information. In other words, coherence can be fully recovered at the level of visibility, whereas sensitivity remains irreversibly limited by attenuation.

We have obtained analytical conditions for the optimal entanglement, the maximal achievable Fisher information, and the boundaries for quantum superiority and quantum advantage, and have characterized their scaling with the photon number. In particular, the two-photon case, which corresponds to the most accessible regime in current photonic experiments, allowed us to map the full $(\alpha,p)$ parameter space, providing an explicit operational diagram that delineates the regions of sub-SQL sensitivity and of practical advantage over single-photon strategies. This highlights, in a quantitative manner, which imperfections act as hard limitations and which can be mitigated by tuning accessible experimental degrees of freedom.

Beyond its relevance for entangled-photon interferometry, our analysis illustrates a more general principle for quantum-enhanced sensing: optimizing interference contrast is not sufficient to optimize information, and loss–imbalance compensation at the level of amplitudes does not translate into an equivalent compensation at the level of metrological gain. Extensions of this framework to include global losses and detector inefficiencies \cite{Oh2017,Datta2011}, as well as mixed or partially distinguishable input states, would further bridge the gap between idealized bounds and realistic implementations, and provide refined benchmarks for near-term quantum sensors.

\section*{Acknowledgment}

This work was supported by the French Defence Innovation Agency (Agence de l’Innovation de Défense, Direction Générale de l’Armement) through the QAFEINE and PARADIS projects (ANR-21-ASTR-007 and ANR-22-ASTR-0027-DA). The authors also acknowledge financial support from the Agence Nationale de la Recherche (ANR) through the EQUINE project (ANR-23-QUAC-0001), and from the French government through its Investments for the Future programme under the Université Côte d’Azur UCA-JEDI project (Quantum@UCA) managed by the ANR (ANR-15-IDEX-01).

\section*{Competing interests}
The authors declare that they have no conflict of interest.

\section*{Data Availability}
The data are available from the authors upon request.


\begin{thebibliography}{99}

\bibitem{dalidet_QA}
R.~Dalidet, A.~Martin, G.~Sauder, S.~Tanzilli, and L.~Labont\'e,
Quantum-enhanced phase sensitivity in an all-fiber Mach-Zehnder interferometer,
arXiv:2602.18354 [quant-ph] (2026).

\bibitem{Barbieri2022}
M.~Barbieri,
Optical Quantum Metrology,
{\em PRX Quantum} \textbf{3}, 010202 (2022).

\bibitem{Paris2009}
M.~G.~A.~Paris,
Quantum Estimation for Quantum Technology,
{\em Int.\ J.\ Quantum Inf.} \textbf{7}, 125--137 (2009).

\bibitem{Lee2002Rosetta}
H.~Lee, P.~Kok, and J.~P.~Dowling,
A quantum Rosetta stone for interferometry,
{\em J.\ Mod.\ Opt.} \textbf{49}, 2325--2338 (2002).

\bibitem{Demkowicz2015}
R.~Demkowicz-Dobrza\'nski, M.~Jarzyna, and J.~Ko{\l}ody\'nski,
Quantum Limits in Optical Interferometry,
{\em Prog.\ Opt.} \textbf{60}, 345--435 (2015).

\bibitem{Rubin2007}
M.~A.~Rubin and S.~Kaushik,
Loss-induced limits to phase measurement precision with maximally entangled states,
{\em Phys.\ Rev.\ A} \textbf{75}, 053805 (2007).

\bibitem{Oh2017}
C.~Oh, S.-Y.~Lee, H.~Nha, and H.~Jeong,
Practical resources and measurements for lossy optical quantum metrology,
{\em Phys.\ Rev.\ A} \textbf{96}, 062304 (2017).

\bibitem{Lee2009Loss}
T.~Lee, S.~Huver, H.~Lee, L.~Kaplan, S.~McCracken, C.~Min, D.~Uskov, C.~Wildfeuer, G.~Veronis, and J.~P.~Dowling,
Optimization of quantum interferometric metrological sensors in the presence of photon loss,
{\em Phys.\ Rev.\ A} \textbf{80}, 063803 (2009).

\bibitem{AtamanMishra2024}
S.~Ataman and K.~K.~Mishra,
Quantum Fisher Information Maximization in an Unbalanced Lossy Interferometer,
{\em Phys.\ Rev.\ A} \textbf{109}, 062605 (2024).

\bibitem{Mitchell2004}
M.~W.~Mitchell, J.~S.~Lundeen, and A.~M.~Steinberg,
Super-resolving phase measurements with a multiphoton entangled state,
{\em Nature} \textbf{429}, 161--164 (2004).

\bibitem{Nagata2007}
T.~Nagata, R.~Okamoto, J.~L.~O'Brien, K.~Sasaki, and S.~Takeuchi,
Beating the Standard Quantum Limit with Four-Entangled Photons,
{\em Science} \textbf{316}, 726--729 (2007).

\bibitem{Afek2010}
I.~Afek, O.~Ambar, and Y.~Silberberg,
High-NOON States by Mixing Quantum and Classical Light,
{\em Science} \textbf{328}, 879--881 (2010).

\bibitem{ThomasPeter2011}
N.~Thomas-Peter, B.~J.~Smith, A.~Datta, L.~Zhang, U.~Dorner, and I.~A.~Walmsley,
Real-World Quantum Sensors: Evaluating Resources for Precision Measurement,
{\em Phys.\ Rev.\ Lett.} \textbf{107}, 113603 (2011).

\bibitem{Qin2023}
J.~Qin, Y.~Deng, H.~Zhong, L.~Peng, H.~Su, Y.~Luo, J.~Xu, D.~Wu, S.~Gong, H.~Liu, H.~Wang, M.~Chen, L.~Li, N.~Liu, C.-Y.~Lu, and J.-W.~Pan,
Unconditional and Robust Quantum Metrological Advantage beyond N00N States,
{\em Phys.\ Rev.\ Lett.} \textbf{130}, 070801 (2023).

\bibitem{Nielsen2023}
J.~S.~Nielsen, J.~S.~Neergaard-Nielsen, T.~Gehring, and U.~L.~Andersen,
Deterministic Quantum Phase Estimation beyond N00N States,
{\em Phys.\ Rev.\ Lett.} \textbf{130}, 123603 (2023).

\bibitem{Huver2008}
S.~D.~Huver, C.~F.~Wildfeuer, and J.~P.~Dowling,
Entangled Fock states for robust quantum optical metrology, imaging, and sensing,
{\em Phys.\ Rev.\ A} \textbf{78}, 063828 (2008).

\bibitem{Hiekkamaki2021}
M.~Hiekkam\"aki, F.~Bouchard, and R.~Fickler,
Photonic Angular Superresolution Using Twisted N00N States,
{\em Phys.\ Rev.\ Lett.} \textbf{127}, 263601 (2021).

\bibitem{Winsten2024}
Y.~Winsten, D.~Cohen, and Y.~Sagi,
Tweezer interferometry with NOON states,
{\em Phys.\ Rev.\ A} \textbf{110}, 043308 (2024).

\bibitem{Demkowicz2009}
R.~Demkowicz-Dobrza\'nski, U.~Dorner, B.~J.~Smith, J.~S.~Lundeen, W.~Wasilewski, K.~Banaszek, and I.~A.~Walmsley,
Quantum phase estimation with lossy interferometers,
{\em Phys.\ Rev.\ A} \textbf{80}, 013825 (2009).

\bibitem{Kolodynski2010}
J.~Ko{\l}ody\'nski and R.~Demkowicz-Dobrza\'nski,
Phase estimation without a priori phase knowledge in the presence of loss,
{\em Phys.\ Rev.\ A} \textbf{82}, 053804 (2010).

\bibitem{Datta2011}
A.~Datta, L.~Zhang, N.~Thomas-Peter, U.~Dorner, B.~J.~Smith, and I.~A.~Walmsley,
Quantum metrology with imperfect states and detectors,
{\em Phys.\ Rev.\ A} \textbf{83}, 063836 (2011).

\bibitem{Dorner2009}
U.~Dorner, R.~Demkowicz-Dobrza\'nski, B.~J.~Smith, J.~S.~Lundeen, W.~Wasilewski, K.~Banaszek, and I.~A.~Walmsley,
Optimal quantum phase estimation,
{\em Phys.\ Rev.\ Lett.} \textbf{102}, 040403 (2009).

\bibitem{Escher2011}
B.~M.~Escher, R.~L.~de~Matos~Filho, and L.~Davidovich,
General framework for estimating the ultimate precision limit in noisy quantum-enhanced metrology,
{\em Nat.\ Phys.} \textbf{7}, 406--411 (2011).

\bibitem{holland_interferometric_1993}
M.~J.~Holland and K.~Burnett,
Interferometric detection of optical phase shifts at the Heisenberg limit,
{\em Phys.\ Rev.\ Lett.} \textbf{71}, 1355--1358 (1993).

\bibitem{Thekkadath2020}
G.~S.~Thekkadath, M.~G.~Mycroft, B.~A.~Bell, C.~G.~Wade, A.~Eckstein, D.~S.~Phillips, R.~B.~Patel, A.~Buraczewski, A.~E.~Lita, T.~Gerrits, S.~W.~Nam, M.~Stobi\'nska, A.~I.~Lvovsky, and I.~A.~Walmsley,
Quantum-enhanced interferometry with large heralded photon-number states,
{\em npj Quantum Inf.} \textbf{6}, 89 (2020).

\bibitem{HuLu2025}
B.-S.~Hu and X.-M.~Lu,
Revisit on Quantum Parameter Estimation Approach for Mach--Zehnder Interferometry,
arXiv:2503.14306 [quant-ph] (2025).

\bibitem{Braunstein1994}
S.~L.~Braunstein and C.~M.~Caves,
Statistical Distance and the Geometry of Quantum States,
{\em Phys.\ Rev.\ Lett.} \textbf{72}, 3439--3443 (1994).

\bibitem{Agarwal2010}
G.~S.~Agarwal, S.~Chaturvedi, and A.~Rai,
Amplification of maximally-path-entangled number states,
{\em Phys.\ Rev.\ A} \textbf{81}, 043843 (2010).

\bibitem{Bezerra2025}
M.~E.~O.~Bezerra, F.~Albarelli, and R.~Demkowicz-Dobrza\'nski,
Simultaneous Optical Phase and Loss Estimation Revisited: Measurement and Probe Incompatibility,
arXiv:2504.02893 [quant-ph] (2025).

\bibitem{HuertaAlderete2022}
C.~Huerta Alderete, M.~Hunter Gordon, F.~Sauvage, \emph{et al.},
Inference-Based Quantum Sensing,
{\em Phys.\ Rev.\ Lett.} \textbf{129}, 190501 (2022).

\bibitem{Wang2024}
K.~Wang, S.~J.~U.~White, A.~Szameit, A.~A.~Sukhorukov, and A.~S.~Solntsev,
Demonstration of lossy linear transformations and two-photon interference on a photonic chip,
{\em Phys.\ Rev.\ Res.} \textbf{6}, 043076 (2024).

\bibitem{Franson1989}
J.~D.~Franson,
Bell inequality for position and time,
{\em Phys.\ Rev.\ Lett.} \textbf{62}, 2205--2208 (1989).

\bibitem{Rarity1990}
J.~G.~Rarity, P.~R.~Tapster, E.~Jakeman, T.~Larchuk, R.~A.~Campos, M.~C.~Teich, and B.~E.~A.~Saleh,
Two-photon interference in a Mach--Zehnder interferometer,
{\em Phys.\ Rev.\ Lett.} \textbf{65}, 1348--1351 (1990).

\end{thebibliography}
\end{document}